\batchmode
\documentstyle[11pt,newpaspb,twoside, epsf]{article}
\begin{document}
\newcommand{\simlt}{\la}
\newcommand{\simgt}{\ga}
\newcommand{\Cfac}{{\cal C}}
\newcommand{\tphici}{\tau_{{}_{\Phi,\rm{c0}}}}
\newcommand{\msol}{M_{\odot}}
\newcommand{\rhon}{\rho_{\rm n}}
\newcommand{\vn}{v_{\rm n}}
\newcommand{\vnr}{v_{{\rm n},r}}
\newcommand{\vi}{v_{\rm i}}
\newcommand{\vir}{v_{{\rm i},r}}
\newcommand{\nn}{n_{\rm n}}
\newcommand{\nnc}{n_{\rm{n,c}}}
\newcommand{\nnci}{n_{\rm{n,c0}}}
\newcommand{\nnic}{n_{\rm{i,c}}}
\newcommand{\nnici}{n_{\rm{i,c0}}}
\newcommand{\cc}{{\rm{cm}^{-3}}}
\newcommand{\xxi}{x_{\rm i}}
\newcommand{\xxic}{x_{\rm{i,c}}}
\newcommand{\xxici}{x_{\rm{i,c0}}}
\newcommand{\PhiB}{\Phi_B}
\newcommand{\tff}{\tau_{\rm{ff}}}
\newcommand{\tni}{\tau_{\rm{ni}}}
\newcommand{\tad}{\tau_{{}_{\rm{AD}}}}
\newcommand{\tadci}{\tau_{{}_{\rm{AD,c0}}}}
\newcommand{\tcore}{t_{\rm{core}}}
\newcommand{\muc}{\mu_{\rm c}}
\newcommand{\muci}{\mu_{\rm{c0}}}
\newcommand{\mdot}{\dot{M}}
\title{The Razor's Edge: Magnetic Fields and Their Fundamental Role
in Star Formation and Observations of Protostellar Cores}
\author{Glenn E. Ciolek}
\affil{New York Center for Studies on the Origins of Life (NSCORT),
Rensselaer Polytechnic Institute, 110 8th Street, Troy, NY 12180,
U.S.A.}
\author{Shantanu Basu}
\affil{Department of Physics and Astronomy, University of Western
Ontario, London, Ontario N6A 3K7, Canada}
\begin{abstract}
We review theoretical models of the early stages of star formation, in
which gravitational collapse is strongly regulated by magnetic fields
and the associated process of ambipolar diffusion. We discuss 
results of numerical simulations and analytical studies of core
formation and collapse, which can be directly tested
against observation. We also focus on recent data which 
are relevant to this theory of star formation, such as: observations of
extended infall in protostellar cores, estimates of evolutionary
timescales $\sim 1$ Myr for cores, measured mass-to-flux ratios of
cores, and the relative alignment of polarization vectors with apparent
cloud elongation. It is shown that in all of these areas, the data
remain compatible with magnetic collapse models which lie within the
observationally allowable range of parameter space. Other areas of
interest (protostellar accretion rates and the presence of core edges)
and issues that remain unresolved or under study (the role of
non-thermal motions and cluster formation) are also discussed. Moreover,
we highlight some differences between our model predictions and those of
highly turbulent star formation models, and discuss how these
differences can be distinguished observationally.
\end{abstract}
%
\vspace{-3ex}
\begin{quote}
``Never multiply explanations or make them more complicated than
necessary. An explanation should be as simple and direct as possible."

``No more things should be presumed than is necessary."
\flushright {---\em William of Occam} (attributed to),
English monk, philosopher Occam's Razor (c. 1285-1349).
\end{quote}
\vspace{-3ex}
\section{Introduction: The Need for a Magnetic Theory of Star Formation}
A robust physical theory is one that explains
observed phenomena and obeys the laws of physics. And, in the sense of
Occam described above, it should be as parsimonious as possible
with the number of assumptions used in its formulation.

There are a number of lines of evidence that strongly suggest that
magnetic fields play a crucial role in a comprehensive theory of star
formation. For instance, observations of magnetic field strengths
by OH Zeeman splitting in dense gas (e.g., Crutcher 1999)
allow a determination of the mass-to-flux ratio $M/\PhiB$ in molecular
clouds. When allowances are made for the random orientations of clouds
and magnetic fields (e.g., see Table 1 of Shu et al. 1999), the
observed values for $M/\PhiB$ in several clouds are found to be
a factor $\sim 2$ below or above the critical value for gravitational
collapse, $(M/\PhiB)_{\rm{crit}}$ (e.g., Mouschovias \& Spitzer 1976).
That the observed values for these clouds are so close to the critical
value indicates that magnetic fields must be instrumental in
their support against self-gravitational collapse. Additionally,
Crutcher's data set also reveals that the magnetic field strength
$B$ scales with the gas density $\rhon$ as $B \propto \rhon^{1/2}$,
consistent with the predictions from models of equilibrium magnetic
clouds (Mouschovias 1976) and ambipolar diffusion models of magnetically
supercritical core formation and collapse in subcritical clouds
(Fiedler \& Mouschovias 1993; Ciolek \& Mouschovias 1994, 1995;
Basu \& Mouschovias 1994, 1995a, b).

Another inextricable link between magnetic fields and star formation
is the angular momentum of stars and protostellar disks.
It has long been known that if absolute conservation of angular momentum
were to hold true throughout the evolution of a cloud, from its
formation to protostellar collapse, then star formation would be halted
by centrifugal forces (e.g., Mestel \& Spitzer 1956). The solution to
this conundrum is magnetic braking of clouds and cores, which allows the
efficient transfer of angular momentum via rotational Alfv\'{e}n waves
along the magnetic field, from a cloud (e.g., Mouschovias \& Paleologou
1979, 1980) or core (e.g., Basu \& Mouschovias 1994, 1995a, b) to an
external medium. Measurements do indeed find that rotational velocities
in clouds are orders of magnitude smaller than that which would occur
under strict conservation of angular momentum (e.g., Goldsmith \&
Arquilla 1985). Observations of near alignment of rotation axes with
polarimetrically determined magnetic field direction in clouds (e.g.,
Heyer et al. 1987) and starless Bok globules (e.g., Kane \& Clemens
1995) also confirm the expected correspondence from magnetic braking
models. Another prediction of these models is retrograde rotation of
clouds and fragments (Mouschovias \& Paleologou 1979), which has also
been seen to occur (Clark \& Johnson 1981; Young et al. 1981).

There are various other pieces of support for the importance of magnetic
fields in star formation, such as the existence of the magnetic flux
problem of star formation (e.g., Mestel \& Spitzer 1956), evidence that 
clouds are in near virial equilibrium, with linewidths that are related
to the Alfv\'{e}n speed (Myers \& Goodman 1988; Mouschovias \& Psaltis
1995), and the potential importance of the Parker instability in
forming molecular cloud complexes (Mouschovias, Shu, \& Woodward 1974).

As argued above, {\em all} of these various observational facts or
threads of reasoning can be explained and tied logically together
through a {\em single} unifying theory of star formation in
magnetically supported clouds. In the next section we summarize results
of ambipolar diffusion models of core and star formation. More
extensive treatments of the history and the relevant physics of the
magnetic models of star formation can be found in the detailed reviews
by Mouschovias (1987, 1996) and Mouschovias \& Ciolek (1999).

\section{Key Features of Ambipolar Diffusion and Core Formation}
\vspace{-2ex}
Ambipolar diffusion, the drift of neutral matter with respect to
plasma and magnetic field, was originally suggested by Mestel \&
Spitzer (1956) as a way that magnetic flux can leak out of interstellar
clouds and induce collapse and fragmentation. A fundamental reevaluation
of this concept was introduced by Mouschovias (1976, 1977, 1978), who
pointed out that the main effect of ambipolar diffusion is to
gravitationally redistribute mass and magnetic flux within the interior
magnetic flux tubes of a cloud, thereby initiating the formation and
collapse of cores within magnetically supported envelopes.
He further argued (Mouschovias 1979, 1982) that core formation will
occur on the ambipolar diffusion timescale, $\tad \simeq \tff^2/\tni$,
where $\tff$ is the free-fall time and $\tni$ is the neutral-ion
collision time in a cloud's inner flux tubes. Early studies
(Mouschovias 1979; Nakano 1979; Lizano \& Shu 1989) focused on 
quasistatic evolution and the approach to core formation.

The general theoretical scenario sketched above was verified by
numerical simulations of the evolution of two-dimensional axisymmetric
(Fiedler \& Mouschovias 1993) and flattened, thin-disk (Ciolek \&
Mouschovias 1994, 1995 [CM94, CM95]; Basu \& Mouschovias 1994, 1995a, b
[BM94, BM95a, b]; Ciolek \& Basu 2000 [CB00]), self-gravitating,
isothermal, magnetic molecular cloud
models. Following the evolution of the central flux tubes of a model
cloud from the initial quasistatic contraction (due to ambipolar
diffusion) to the later phase of dynamical collapse, from initial
densities $\nnci \sim 10^3~\cc$ to
final densities $\nnc \sim 10^{11}~\cc$, these model simulations found
that supercritical [$(M/\PhiB)_{\rm{core}} \geq (M/\PhiB)_{\rm{crit}}$]
cores did indeed form and separate within massive,
subcritical [$(M/\PhiB)_{\rm{env}} < (M/\PhiB)_{\rm{crit}}$]
envelopes. The core masses found in these models were typically
in the range $\sim 3-30~\msol$, with characteristic radii in the range
$0.1 - 0.3$ pc, depending on the choice of model parameters such as the
initial central mass-to-flux ratio $\muci$ (in units of the critical
value for collapse) and the initial central degree of ionization
$\xxici$ ($=\nnici/\nnci$, where $\nnici$ is the ion density).
Mean core densities $\langle \nn \rangle$ were $\sim 10^4~\cc$.
A magnetic disk-model was presented by Crutcher et al. (1994) to
explain the evolution of the Barnard 1 (B1) molecular cloud. Model
predictions for
the B1 core were found to be in excellent agreement with observations;
predicted values of the core mass, and mean density and magnetic field
strength were within 10\% of the observed values. Further discussion
of quantitative comparison of ambipolar diffusion models with
observations can be found in \S~3.1 below, as well as the paper by
Bacmann in this volume.   

The disk-model clouds also included other interesting physical effects.
Dynamical effects of dust grains (charged and neutral) were
studied in detail by Ciolek \& Mouschovias (1993, 1994, 1995). Enhanced
friction due to collisions with grains increases the coupling of neutral
particles to the magnetic field, and can significantly lengthen the time
required to form a core. Another interesting result from these models is
that ambipolar diffusion reduces the dust-to-gas ratio in cores (also
discussed in Ciolek \& Mouschovias 1996), which, in turn, affects the
ion chemistry in contracting cores, since ion and electron abundances
in the gas phase are dependent on the rate of capture on grains (an
analytical derivation is provided by Ciolek \& Mouschovias 1998).
Rotation and magnetic braking were incorporated by Basu \& Mouschovias
(1994, 1995a, b). These studies showed that magnetic braking is
generally very efficient (occurring on a timescale much shorter than
$\tad$), allowing a core to form and collapse with very little hindrance
by centrifugal forces; the angular momentum per unit mass $J/M$ in
these models was found to consistent with observations of rotating cores
and estimates of likely values in the early solar nebula. The effect of
the interstellar ultraviolet radiation field was investigated by Ciolek
\& Mouschovias (1995). They found that the increased ionization due to
an external UV field greatly decreased the efficiency of ambipolar
diffusion in cloud envelopes, significantly reducing the mass infall
rate beyond the boundary of a supercritical core.

Further details of the models can be found in the studies cited above,
as well as the more recent reviews mentioned in \S~1. Some more
discussion of the L1544 model of CB00 is provided in \S~3.1 below.
\vspace{-2ex}
\section{Controversies, Unresolved Questions, and Work in Progress}
\vspace{-1ex}
\vspace{-1ex}
\subsection{Extended Infall in Cores}
The protostellar core within the L1544 cloud, located in Taurus, has
received considerable attention in recent years by various workers.
One such study was by Tafalla
et al. (1998), who determined that the neutral gas in the core was
infalling at a speed $|\vn| \sim 0.1~{\rm{km}}~{\rm s}^{-1}$ on a scale
$\sim 0.1~\rm{pc}$. A later study by Williams et al. (1999) used
$\rm{N_{2}H^{+}}$ lines to find the infall speeds of the ionized gas
component, and found $|\vi| \sim 0.08~{\rm{km}}~{\rm s}^{-1}$ on
smaller scales $\sim 0.02~\rm{pc}$. Both papers contrasted their
findings with the ambipolar diffusion models of CM95 and BM94, and
suggested that a discrepancy
existed because certain ambipolar diffusion models did not 
exhibit such large infall speeds at extended radii. However,
as noted by CB00, many of the features observed
in L1544 (such as the ion infall speed and density structure on smaller
scales) were in fact consistent
with the previously published generic ambipolar diffusion models,
if one considers the observed region in L1544 to lie {\em within} a 
supercritical core. Even in the absence of a specific model, magnetic
restoring forces must be important in the evolution of L1544, since
the magnitude of the inferred infall speeds are {\em less} than
characteristic speeds such as the isothermal sound speed $C$ in the gas
($\simeq 0.2~\rm{km}~\rm{s}^{-1}$ for this cloud) and the
free-fall speed at these radii ($\gg C$). And, as the linewidths in the
L1544 core are very narrow (Williams et al. 1999), ``turbulent" support
cannot be invoked to explain the evidently retarded collapse.

CB00 presented
an ambipolar diffusion  model specifically designed to study the
evolution of the L1544 core. They modeled L1544 as a disk-like cloud
inclined at an angle $\theta = 16^\circ$ to the plane of the sky; the
resulting model evolution produced a core with radius 0.3 pc,
which is indeed greater than the spatial scale studied by
Tafalla et al. (1998) and Williams et al. (1999).
Figure 1 ({\it left}) shows the spatial profile of the density at ten
different times $t_{j}$ (see caption). Also shown ({\it right}) is the
infall speed of the neutral gas $\vn$ ({\it solid}
lines) and the ions $\vi$ ({\it dashed} lines) at the same times
$t_{j}$. Qualitatively the overall behavior is similar to that seen
in the earlier published ambipolar diffusion models, including the
flattening of the density profiles at small radii, which is very
similar to the density profiles seen in sub-mm dust continuum studies of
prestellar cores (e.g., Andr\'{e}, Ward-Thompson, \& Motte 1996). 
In Table 1, CB00 model predictions at times $t_2$ and $t_3$ are
compared with
various observed quantities (such as the mass $M$, density $\nn$, radial
neutral and ion speeds $\vnr$ and $\vir$, and the mean line-of-sight
magnetic field strength $\langle B \rangle_{\rm{los}}$) at several
different radial locations. Given the likely observational
uncertainties, we note that {\em all} of the model predictions are in
reasonable agreement with the actual measured values --- including the
extended infall observations and the measurement of 
$\langle B \rangle_{\rm{los}}$ by Crutcher \& Troland (2000), made 
subsequent to our prediction.
\footnote{The values for the drift speed $v_{{}_{\rm D}}$
($= \vi - \vn$) which lie in the range $0.02~\rm{km}~\rm{s}^{-1}$ to
$0.04~\rm{km}~\rm{s}^{-1}$ for the L1544 model, are also in agreement
with the limits deduced by Benson et al. (1998),
who inferred $v_{{}_{\rm D}} \simlt 0.03~\rm{km}~{\rm s}^{-1}$ from
velocity differences of neutral molecules (CCS and $\rm{C_{3}H_{2}}$)
and $\rm{NH_{3}}^{+}$ in sixty dense cores.}
This suggests that the current state $t$ of L1544 is described by the
theoretical model for $t_{2} \leq t \leq t_{3}$.

The CB00 L1544 model is thus able to {\em simultaneously}
explain the observations of several {\em disparate} observations at many
{\em different} radial locations. We also note that the magnetic
models were fully developed prior to 1995, and were applied, without
modification, to the recent observations of L1544 in a straightforward
fashion. That a {\em single} magnetic model can explain the observations
with (e.g., L1544) and without extended infall through variations
of input parameters within the observationally allowable range, is
particularly appealing, and in line with our discussion in \S~1 of what
attributes (simplicity, elegance, self-consistency, and predictive
power) a proper physical theory should have.
\begin{figure}
\vspace{-3ex}
\plotone{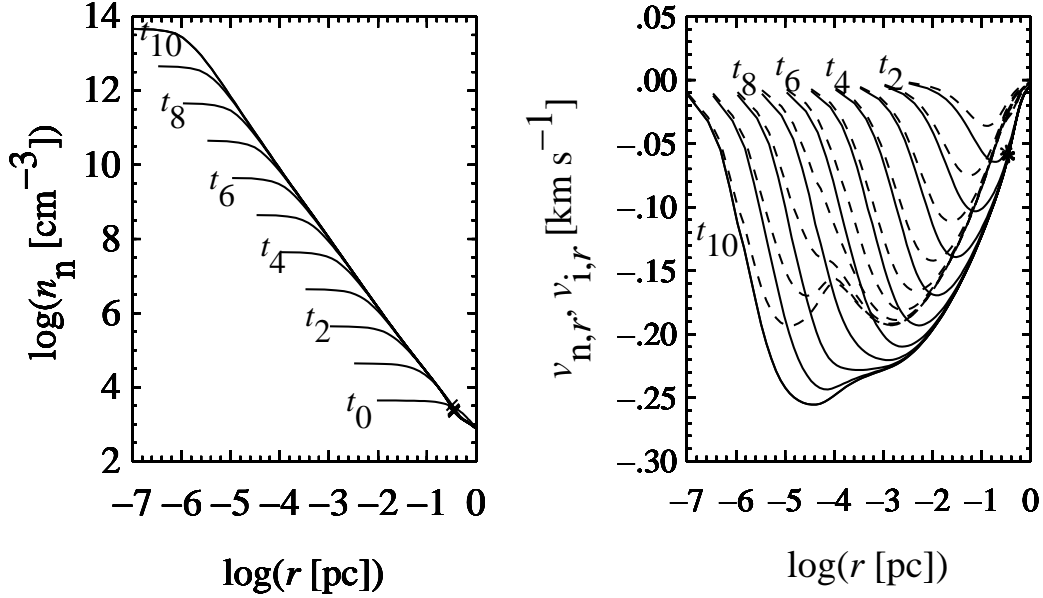}
\vspace{-3ex}
\caption{Spatial profiles of physical quantities in the L1544 model
cloud of CB00, at ten different times
$t_{j}$ ($j$=0, 1, 2,...,10), at which the central density
at time $t_j$ is a factor $10^j$ greater than the initial central
density. These times are, respectively, 0, 2.27, 2.60, 2.66, 2.680,
2.684, 2.685, 2.6856, 2.68574, 2.68577, and 2.68578 Myr. An
asterisk on a curve (present for $t_{j} > t_{0}$), locates the
instantaneous radius of the critical magnetic flux tube. Left:
density. Right: infall speed of the neutrals ({\it solid} curves)
and the ions ({\it dashed} curves).}
\end{figure}
\vspace{-3ex}
\begin{table}
\begin{center}
TABLE 1 \\
{\sc Physical Quantities in the L1544 Core \\}
\begin{tabular}{lllll}
\hline
\hline
\mbox{\hspace{6em}} & CB00: $t_{2}$ & Observed  & CB00: $t_{3}$\\
\hline
$r \simeq$ 0.14 pc : ${}^{\rm a}$ &    & \\
\hspace{1em} $M$ &  $7.1~\msol$ & $8~\msol$ & $7.7~\msol$ \\
\hspace{1em} $|\vnr|$ & $0.09~\rm{km}~\rm{s}^{-1}$ & $0.1~\rm{km}~{\rm s}^{-1}$ & $0.11~\rm{km}~\rm{s}^{-1}$ \\
\hspace{1em} $|\vir|$ & $0.06~\rm{km}~{\rm s}^{-1}$ & ------  & $0.07~\rm{km}~\rm{s}^{-1}$\\
$r \simeq$ 0.06 pc : ${}^{\rm b}$ & & \\
\hspace{1em} $\langle B \rangle_{\rm{los}}$   & $13.4~\mu$G & $11~\mu$G & $15.5~\mu$G \\ 
$r \simeq$ 0.02 pc : ${}^{\rm c}$ & & \\
\hspace{1em} $|\vnr|$ & $0.07~\rm{km}~\rm{s}^{-1}$ & ------ & $0.14~\rm{km}~\rm{s}^{-1}$ \\
\hspace{1em} $|\vir|$ & $0.05~\rm{km}~{\rm s}^{-1}$ & $0.08~\rm{km}~{\rm s}^{-1}$ & $0.11~\rm{km}~{\rm s}^{-1}$ \\
\hspace{1em} $\nn$ & $2.3 \times 10^5~\cc$ & $4 \times 10^5~\cc$ & $3.3 \times 10^5~\cc$ \\
$r \simeq 1.2 \times 10^3~\rm{AU}$ : ${}^{\rm d}$ &    & \\
\hspace{1em} $\nn$ &  $4.1\times 10^{5}~\cc$ & $7 \times 10^{5}~\cc$ & $2.2\times 10^{6}~\cc$ \\ 
\hline
\end{tabular}
\end{center}
${}^{\rm a}$ Observed values taken from Tafalla et al.
(1998). ${}^{\rm b}$ Observed values taken from Williams et al. (1999).
${}^{\rm c}$ Observed value taken from Crutcher \& Troland (2000),
measured after the prediction made by CB00.
${}^{\rm d}$ Observed value taken from Ward-Thompson et al. (1999)
and Bacmann et al. (2000).
\end{table}
\begin{figure}
\vspace{-35ex}
\plotone{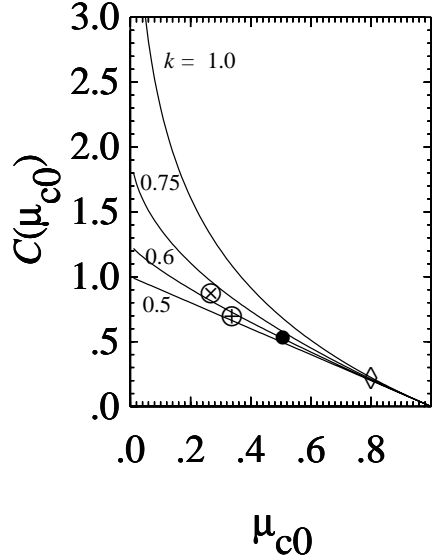}
\vspace{-25ex}
\caption{Initial flux constant $\Cfac(\muci) = \tcore/\tphici$ for
several model clouds with $k =0.5$,
0.6, 0.75, and 1.0, respectively. Also shown are the values found
from previously published numerical simulations: model A of CM94 
(located by the center of $\otimes$), model 2 of
BM94 ($\oplus$), model 5 of BM95b ($\bullet$), and the
L1544 model cloud of CB00 ($\diamond$).
Above $\muci \simgt 0.5$, $\Cfac(\muci)$ is significantly less than
unity.
\vspace{-10ex}
}
\end{figure}
\vspace{-2ex}
\subsection{Timescale for Star Formation}
From the observed infall data discussed above for L1544, one may deduce
a kinematical timescale $\approx r/|\vn| \sim 1$ Myr. Recent
statistical studies of protostellar cores and young stellar objects
(Lee \& Myers 1999; Jijina, Myers, \& Adams 1999) have also estimated
a similar evolutionary timescale for star formation. These studies
also suggested that the observed timescales were in disagreement
with earlier published ambipolar diffusion models (e.g., CM95;
BM94), which had cores
forming on a timescale $\tcore \simeq \tphici \sim 10 \tff$, where
$\tphici = \tadci/2 = (\tff^2/2 \tni)_{\rm{c0}}$ is the initial central
flux-loss timescale, as proposed early on by Mouschovias (see \S~1).

However, the relation $\tcore \simeq \tphici$ is strictly valid only
for highly subcritical model clouds, i.e., clouds with $\muci \ll 1$.
In fact, ambipolar diffusion models with less subcritical initial
mass-to-flux ratios have $\tcore \ll \tphici$.
For instance, the L1544 model of CB00 had $\muci = 0.8$ and
$\tcore = 0.2 \tphici$; numerically this was equal to 1.3 Myr,
similar to the values from the statistical studies cited above.
Hence, $\tcore$ is dependent on $\muci$. Ciolek \& Basu (2001; [CB01])
revisited the problem of core formation by ambipolar diffusion, and
derived an analytical expression for $\tcore$ valid for {\em all}
subcritical clouds ($\muci \leq 1$). Analyzing the evolution during
the epoch of core formation, they found that
$\tcore = \Cfac(\muci)\tphici$, where $\Cfac(\muci)$ is the {\em initial
flux constant} of a cloud, and is a measure of ``how far away" 
a cloud is from the critical mass-to-flux ratio (i.e., $\muc = 1$ in
dimensionless units).
For clouds in which neutral-ion collisions
are predominant, and approximating the relation between ions and
neutrals in the central flux tubes of a cloud by $\nnic \propto \nnc^k$
($0.5 \simlt k \simlt 1$, Ciolek \& Mouschovias 1998; [CM98]), 
 \begin{eqnarray}
 \label{Cfaceqa}
 \Cfac(\muci)  &=& \frac{1 - \muci^{2(1-k)}}{2\left(1-k\right)} 
 \hspace{4em}{\rm{for}}~k < 1, \\
 \label{Cfaceqb}
 &=& \ln\left(\frac{1}{\muci}\right) \hspace{5em}{\rm{for}}~ k=1
 \end{eqnarray}
(a derivation is given in \S~2.1 of CB01).

Figure 2 shows $\Cfac(\muci)$ for several model clouds with different
values of $k$ (see caption). Also shown are the values of
$\Cfac(\muci) = \tcore/\tphici$ for previously published ambipolar
diffusion models; the results from the numerical simulations are in
excellent agreement with the analytical expressions (\ref{Cfaceqa}) and
(\ref{Cfaceqb}). For most realistic values of $k$, clouds with
$\muci \ll 1$ have $\Cfac(\muci) \sim 1$, and, therefore,
$\tcore \approx \tphici$, as Mouschovias (1979) originally argued.
However, for $\muci \simgt 0.5$, $\Cfac(\muci) \ll 1$, with 
$\Cfac(\muci) \simeq (1-\muci)$ for $1 - \muci \ll 1$. This region of
parameter space is especially of interest, since, as noted in \S~1,
Crutcher (1999) has measured mass-to-flux ratios in several clouds
consistent with $\muc \simgt 0.5$. While it is likely that Crutcher's
data set contains clouds that are evolving towards or have recently
exceeded the critical state, and are therefore beginning to collapse
(which is what we suggest has occurred in L1544; see \S~3.1 above), it
may also contain typical values of $\muci$. {\em This means that
$\tcore$ will be $\ll \tphici$ for clouds with these particular values
of $\muci$.} Numerically,
$\tphici \approx 7 \left(\xxici/10^{-7}\right)~\rm{Myr}$,
\footnote{Note that $\xxici$ can be as small as $10^{-8}$ (CM98) so that
a low $\xxici$ is another means of reducing the flux-loss timescale
(e.g., see models 8 and 9 of BM95a), and, hence, $\tcore$.}
where $\xxici = \nnici/\nnci$ has been normalized to values typical of
the interior of molecular clouds for
$10^{3}~\cc \simlt \nnc \simlt 10^{4}~\cc$ (e.g, CM98). It follows
then that $\tcore = \Cfac(\muci)\tphici \simlt 4$ Myr for $\muci > 0.5$.
{\em Thus, the recent statistical analyses suggesting that star
formation occurs in some regions on timescales $\sim$ 1 Myr may be
explained by observations of subcritical clouds and ambipolar diffusion
models with central mass-to-flux ratios slightly below the critical
value for collapse.} Further discussion of the consistency and
implications of ambipolar diffusion models with estimated timescales
for protostellar evolution, including the effect of dust grains, is
presented in CB01.
\vspace{-1ex}
\subsection{Core ``Edges"}
Bacmann et al. (2000) have recently conducted an extensive 
ISOCAM absorption survey of prestellar cores, including L1544, and made
quantitative comparison of their results to theoretical models of core
formation and collapse. For the most part, they found that the
ambipolar diffusion
models of CM95 and BM94 provided the best model fits to their data
in the interiors of cores. However, in some cores they found that on
larger scales the cores effectively had ``edges", where the density
decreased rapidly with increasing radius --- much more so than
predicted by the theoretical models. For L1544,
$r_{\rm{edge}} \sim 0.14$ pc.
\footnote{This value is after deprojection. The measured value for
$r_{\rm{edge}}$ quoted by Bacmann et al. (2000) is $\sim 0.04$ pc.} 
Bacmann et al. suggested that
the presence of these edges indicates that the current theoretical
models neglect some important physics on larger scales, such as
turbulent\footnote{In this paper, we use the term ``turbulent'' quite generally,
to refer to all forms of non-thermal motions, whether an ensemble of
wave motions or a true turbulent cascade.} support, which has not yet 
been incorporated into the ambipolar diffusion simulations. However, if the
turbulence were due to hydromagnetic waves (HMW), as advocated by
Mouschovias (1987) and Mouschovias \& Psaltis (1995), then it is
possible that the reason the non-turbulent CB00
model matches observations for $ r \simlt r_{\rm{edge}}$
in L1544 is due to the {\em decay} of the turbulence for
scales less than the Alfv\'{e}n minimum lengthscale
$\lambda_{\rm A} = \pi v_{\rm A} \tni$
(Kulsrud \& Pearce 1969); $v_{\rm A} = B/(4 \pi \rhon)^{1/2}$
is the Alfv\'{e}n speed in the neutral gas. As first proposed by
Mouschovias (1987), ambipolar diffusion will damp HMW for scales
$< \lambda_{\rm{A}}$, resulting in a loss of support interior to this
region when the HMW turbulence decays away. Formation and collapse of
cores can then proceed via the `standard' self-initiated ambipolar
diffusion models, described earlier. Numerically, the expression for
the damping length
$\lambda_{\rm{A}}= 0.41 (B/10\mu{\rm G})(10^3\cc/\nn)^{3/2}(10^{-7}/\xxi)~\rm{pc}$,
which, when evaluated for the initial parameters of the CB00 L1544 model, is
$\simeq 0.1$ pc. Hence, if HMW were to be important in L1544,
we would expect that a non-turbulent ambipolar diffusion model would
deviate from the actual evolution on scales $\simgt \lambda_{\rm{A}}
\sim r_{\rm{edge}}$, as observed. Moreover,
if decay of HMW has occurred for scales $ < \lambda_{\rm{A}}$, the
linewidths should be narrow in the core, which is again observed in
L1544.
\vspace{-2ex}
\subsection{Protostellar Accretion Rates}
The later stages of core collapse has also been investigated.
Basu (1997) studied semi-analytically the approach of collapsing
cores to higher densities than studied in the earlier numerical
simulations. He suggested that ambipolar diffusion would continue to
affect the increasingly dynamical infall even during the higher density
stages of evolution, and also that the accretion rate $\mdot$ would
become significantly larger than the well-known, time-independent
expression
$\mdot \simeq C^3/G = 1.6\, (T/10{\rm{K}})^{3/2}~\msol~\rm{Myr}^{-1}$
found in self-similar collapse models of a singular isothermal sphere
(Shu 1977). This was confirmed by
Ciolek \& K\"{o}nigl (1998; [CK98]), who followed
the collapse of a magnetic protostellar core (including the effect of
ambipolar diffusion) up to and beyond the formation of a finite mass
object at $r=0$, which they dubbed ``point-mass formation" (PMF). They
followed the evolution in their models up to the time at which a
$1~\msol$ object had accumulated at the center of a core. The accretion
rate in their typical model reached a maximum value of
$9.4~\msol~\rm{Myr}^{-1}$ near the onset of PMF, and then
declined to approximately half that value by the time a $1~\msol$
protostar had formed. During the later stages of the evolution of their
model, ambipolar diffusion affected the accretion rate through the
formation of a hydromagnetic shock at radii $r \simgt 100~\rm{AU}$;
the relevant physics of the inflow and accretion of the numerical model
was also captured in a companion self-similar collapse study presented
by Contopoulos, Ciolek, \& K\"{o}nigl (1998).

The duration of the CK98 simulations spanned the evolution of a
protostar from a class 0 to a class I object. They suggested that the
time-dependent accretion occurring in their models was in accordance
with the analysis of Bontemps et al. (1996), who proposed that a decline
in the observed momentum flux of CO outflows from young stellar objects
with decreasing envelope mass indicated a time-dependent $\mdot$ during
this transitional phase of a protostar's life. CK98 also compared
their results to the measurements of the gas in the infalling disks
of HL Tau (Hayashi, Ohashi, \& Miyama 1993) and L1551-IRS5 (Ohashi et al.
1996). The late-time accretion rate and gas mass of the CK98 model at
$r \simgt  500~\rm{AU}$ were consistent with observed values for both
objects.

\subsection{Magnetic Field Strengths and Alignment}

A remarkable feature of the magnetic field strength measurements
compiled by Crutcher (1999) is that they are consistent with the
intrinsic mass-to-flux ratios (in units of the critical value) of all
cores being very close to unity (see earlier discussion in 
\S\ 1). Additionally, the observed correlation $B \propto \rhon^{1/2}$, 
which implies some degree of flattening along the mean magnetic field
direction, can even be improved if one accounts for internal support
due to non-thermal motions (Basu 2000). {\em The above two features of
the measurements implies that cores have dynamically important magnetic
fields and are characterized by force balance along magnetic field
lines.} 

Recent submillimeter polarization maps of dense regions of molecular
clouds (Matthews \& Wilson 2000; Ward-Thompson et al. 2000), in which
the polarization vectors truly sample the dense gas, reveal that the
magnetic field direction is invariably either parallel to the apparent
minor axis of the objects or has a small offset angle $\Psi$ relative
to it. Ward-Thompson et al. (2000) measured $\Psi \simeq 30^{\circ}$
for three cores (including L1544) and wondered how strong magnetic
fields were compatible with any nonzero $\Psi$. The resolution to this
issue comes from the fact that the axisymmetric assumption in the
magnetic models must be relaxed, even while the assumption of a
dynamically important magnetic field is maintained. As pointed out by
Basu (2000), any non-axisymmetry in the core shape (which makes them
triaxial bodies) results in a finite probability of viewing a nonzero
$\Psi$. Figure 3 illustrates how a single object can appear to have any
possible value of $\Psi$, depending on the viewing angles
($\theta, \phi$) defined on a spherical viewing surface around the
triaxial body. However, if the shortest axis of the body is along the 
mean magnetic field direction, there is an expected correlation towards
$\Psi=0$ for a large sample of measurements (Basu 2000). Hence, large
data sets can easily distinguish between this scenario and that
advocated by Ballesteros-Paredes, Vazquez-Semadeni, \& Scalo (1999),
who use a highly turbulent model (with a weak magnetic field) to claim
there should be no correlation between cloud elongation and magnetic
field direction.

The above explanation for the offset $\Psi = 29^{\circ}$ observed in
L1544 by Ward-Thompson et al. (2000) means that the axisymmetric
assumption employed by CB00 needs to be relaxed, if only slightly.
{\em The correspondence of the Crutcher \& Troland (2000) field strength
measurement with the prediction of the CB00 model, and the
overwhelming evidence for flattening along the mean field direction in
a variety of cores ($B \propto \rhon^{1/2}$) implies that the oblate
geometry of the CB00 model is a reasonable approximation.}

\begin{figure}
\vspace{-3ex}
\plotone{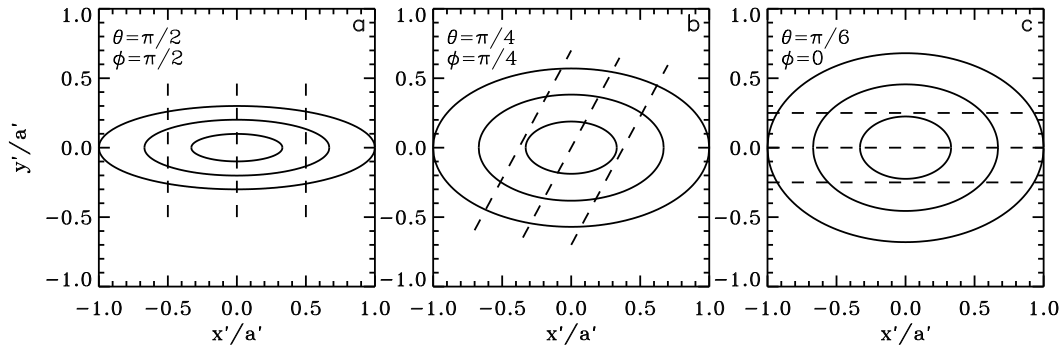}
\vspace{-3ex}
\caption{
Simulated contours (solid lines) and mean polarization direction
(dashed lines) for a triaxial
body with axial ratios $\xi = c/a = 0.3$ and $\zeta = b/a =0.6$ seen from three
sets of viewing angles $(\theta, \phi)$. The
The $x'$ and $y'$ axes are in the plane of the sky and are chosen to lie
along the projected major and minor axes, and $a'$ is the apparent
semi-major axis. The dashed lines
lie along the projected direction of the intrinsic $z$-axis.
a) $(\theta, \phi) = (\pi/2,0)$, yielding
an apparent axis ratio $q = 0.3$ and offset angle $\Psi = 0^{\circ}$.
b) $(\theta, \phi) = (\pi/4,\pi/4)$, yielding $q = 0.57$ and
$\Psi = 28^{\circ}$.
c) $(\theta, \phi) = (\pi/6,0)$, yielding $q = 0.68$ and
$\Psi = 90^{\circ}$.
}
\end{figure}
\subsection{Turbulence and Cluster Formation}
Individual regions which form stars likely must damp their internal
turbulent motions and also overcome the support due to the mean magnetic 
field.
Since these motions may damp preferentially at high density or column density
where the coupling of charged particles is reduced, and the mean magnetic 
field effectively introduces a threshold column density for collapse (\S\ 2), 
a fruitful approach to understanding cluster formation may be to study 
the properties of peaks above a critical threshold in a turbulent column 
density field.
Basu \& Pudritz (2000) have studied the properties of a Gaussian field
of column density fluctuations and how it relates to the properties of
observed clusters. As they point out, star formation in terms of thresholds, 
rather than say, a preferred scale of fragmentation, is qualitatively in better 
agreement with observations. The threshold idea leads to the result
that regions of high mean column density (close to a critical threshold) 
are likely
to form stars (in regions exceeding the threshold) that are more closely 
spaced and typically more massive than in regions of low mean column density
(far from the threshold). In linear fragmentation theory, regions of 
high column density yield a smaller fragmentation scale as well, but also 
favor stars of relatively lower masses.

Finally, the pressure due to an external column of (turbulent) matter
may play a role in limiting the size of infall regions and creating
core ``edges'' at smaller radii than in models without external turbulent 
motions (see related discussion in \S\ 3.3).
Despite objections to the concept of a confining pressure raised by
Ballesteros-Paredes et al. (1999) on the basis of a
single simulation, we note that an external pressure 
must exist simply due to the {\it weight} of an external cloud or 
cloud complex. 
While turbulent motions obviously exert kinetic effects on some scale,  
neglecting external pressure is as absurd as neglecting the Earth's 
atmospheric pressure on a windy day!
\vspace{-3ex}
\section{Summary}
In this paper we have described several aspects of the theory of
star formation in
magnetic interstellar clouds, first formulated in the 1970's and
developed further in the past decade. This single theoretical framework 
is consistent with (and can predict)
various seemingly disparate facts such as the
inefficiency of star formation, the structure of regions of core formation
and collapse, magnetic field strengths and mass-to-flux
ratios, transport of angular momentum, time-dependent mass accretion rates,
magnetic field orientations in clouds, etc.
In fact, specific models for core formation and evolution
in two molecular
clouds (Barnard 1 and L1544) have already been presented, and found to
be in good agreement with observations for these two clouds. We have
also discussed how recent data related to extended infall, timescales
for star formation, and the alignment of magnetic fields in cores, are
compatible with the results of magnetic, ambipolar diffusion models.
Our approach is to make quantitative predictions and, where possible, 
clarify differences with other competing scenarios. For instance,
sub-mm polarimetric observations indicating some degree of alignment of
magnetic fields and core axes strongly favor magnetic, ambipolar
diffusion models, rather than highly turbulent star formation models
advocated by some other workers, which predict no such correlation.
It is through such rigorous testing that a scientifically valid theory
of star formation will ultimately be determined.

\acknowledgements We would like to thank the conference organizers for
their generosity and warm hospitality. GC's research is supported by
the NY Center for Studies on the Origins of Life (NSCORT) at RPI under
NASA grant NAG5-7598; his travel funds were provided by an American
Astronomical Society International Travel Grant.
SB is supported by a grant from the Natural Sciences and Engineering
Research Council of Canada.

\end{document}